# Adaptive Control of Surge Impedance for Electric Motors in Motor Drive Systems

Fahmid Sadeque, Fariba Fateh, JiangBiao He, and Behrooz Mirafzal

*Abstract* — This article studies the possibility of controlling the surge impedance of the electric motor in motor drives. The existing solution to suppress (or eliminate) the reflected wave impact on the motor insulation run by a Si-IGBT or SiC-MOSFET-based drive is to use either a sinewave filter or $dv/dt$ filter. The alternative solution suggested in this article is to implement a high-bandwidth electronic circuit at the end of the cable or at the motor terminals to match the surge impedance of the cable and motor. The high-frequency voltage ringing due to the reflected waves in motor drives is around 1 $MHz$, depending on the cable parameters and the length of the cable. In the proposed method, the electronic circuit can quickly detect the $dv/dt$ rise and fall edges and adjust the electronic circuit equivalent impedance when pulses arrive the motor terminals. Thus, the cable and motor surge impedances can be matched over a short time to prevent reflected waves. As a result, the leakage currents passing through the ball bearing and overvoltage stress on the motor insulation can be suppressed significantly.

## I. INTRODUCTION

Higher efficiency, reliability, and power density are the desired features for all electric devices and systems, such as powertrains in more electric ships and aircraft, and electric vehicles. These features can be enhanced using fast high-voltage semiconductor switches, making the powertrain of propulsion systems more compact [1]-[3]. The use of high-speed generators and motors lead to the utilization of ultrafast active rectifiers and inverters [4]-[14]. However, the faster switches operating at higher voltage levels can lead to new technical challenges for electric machine designers in propulsion systems [15]. The high voltage rate creates traveling waves in the cable between the drive and motor, and thus, generates high-frequency voltage ringing across the stator windings with high-voltage spikes. The voltage stress has negative effects on the motor insulation and ball bearings and can shorten the motor lifetime [16]-[21]. Though the reflected waves phenomenon in motor drives has been well studied, the trend toward using wide bandgap (WBG) switches, such as Silicon Carbide (SiC) power MOSFETs, in propulsion systems requires novel solutions to mitigate the high-frequency high-voltage stress that might degrade the motor lifetime. The voltage stress on motor insolation can result in incipient faults, a propulsion system's failure, and extra expenses associated with unexpected maintenance [22]-[28]. Therefore, many methods have been reported on the fault tolerance techniques for inverters and motors, e.g., [29]-[32].

The problems of leakage currents through ball bearing and voltage stress on the stator insolation of ac motors can be mitigated using sine-wave or $dv/dt$ filters. This problem exits in ac motors fed by drives, for both induction and synchronous (permanent magnet motors), or new designs of less rare-earth permanent magnet (PM) motors [33]. While reflected wave phenomenon exists when the cable between the inverter (drive) and motor is relatively long, the high-frequency common-mode and differential-mode issues can shorten the lifetime ac motors and interfere with electronic circuits used in EVs, aircraft electric systems, and assembly lines [34]-[35]. Inverters play the main role in the drives to control the speed or position of ac motors, and in power systems to control the injected power to the grid in grid-following mode and to control the voltage and frequency of an islanded microgrid in grid-forming mode [36]-[47]. Inverters in power grid if equipped with WBG switches can also shorten the lifetime of the inverter-side inductor of an LC or LCL filter due to high $dv/dt$ switching transients.

There are many studies about high-frequency models of cables and ac motors to study reflected waves, voltage stress on motor insulations, and ball bearing leakage currents, e.g. [48]-[51]. Also, many techniques have been presented to mitigate high surge voltages. One of the main technologies used in the industry is to install a passive LC or RLC filter at the drive terminals to attenuate the $dv/dt$ to an acceptable level [52], [53]. This solution is straightforward and practical, but it adds significant size, weight, and losses to propulsion systems. Particularly, if the drive is configured with WBG switches and switched at high frequency, the filter loss will be dramatic, degrading the overall efficiency and requiring more cooling capacity. Eventually, meeting the targets of system efficiency and power density could be challenging. The second technology for of $dv/dt$ mitigation is to implement an active filter to reduce the $dv/dt$ level. In [54], an active reflected wave canceller circuit consisting of two additional active switches, two capacitors, and two diodes is connected at the output of each inverter phase leg in order to break each rising/falling edge of the output voltage by two steps. It shows that such an active filter solution has lower losses and lower physical dimensions, but the additional gate driver circuits, the required capacitors and phase inductors, and the high-bandwidth edge detection requirements may increase the cost and control complexity of the propulsion systems. Another low-loss active filter is introduced to clamp the maximum motor terminal voltage and alter the rising/falling edge of the voltage pulses [55]. Nevertheless, RLC passive power components are still required in this solution, in addition to voltage sensors and active switches, making the system much more complicated. An easy solution to address the surge voltage stress might be to oversize the insulation strength of the motor stator windings, such as thickening the resin or varnish on the windings [56]. However, this solution increases the motor fabrication cost and reduces motor power density. Therefore, a better solution, which on the one hand, will not reduce the power density and efficiency of the drives, and on the other hand, will mitigate the high voltage spikes on the motor winding terminals, will be of paramount necessity.

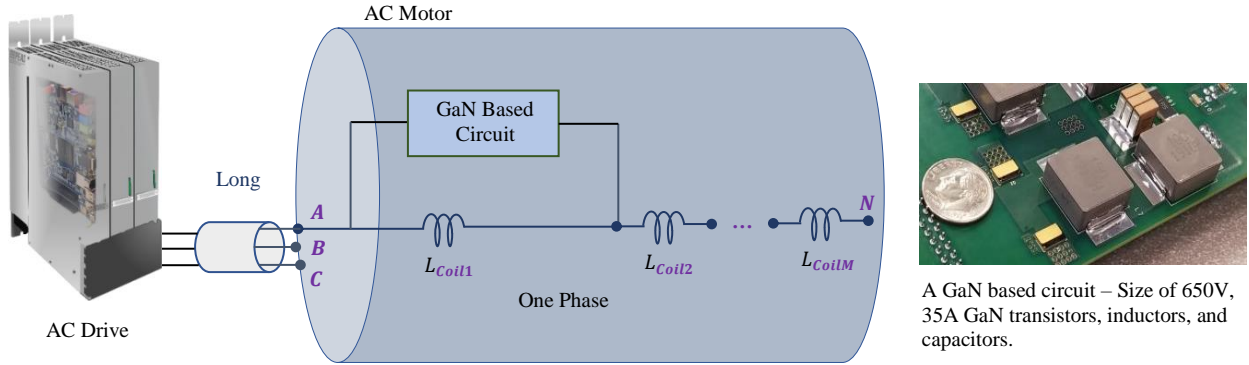

Fig. 1. Adaptive surge impedance circuit implemented on the first coil in each phase.

This article proposes an adaptive surge impedance solution to enhance motor performance in modern motor drive systems. The proposed technology adaptively matches the cable and motor surge impedances. In the proposed solution, the surge impedance of stator windings in an ac motor is changed by implementing a variable impedance branch. The value of this technology is to eliminate the need for any bulky sinewave filter. This technique can extend the motor lifetime and indirectly enhancing motor drive power density and efficiency using WBG switches, which can significantly impact more electric aircraft and shipboards, where power density, efficiency, and reliability are the critical design factors.

## II. SYSTEM DESCRIPTION

This section provides a brief description of the system elements. The controller proposed in this article reforms the conventional surge voltage mitigation approaches from using bulky passive filters to a small-size GaN-based adaptive impedance matching circuit. Fig. 1 illustrates the schematic of the system for the proposed technique for surge impedance control. In motor drive systems, an electric motor is run by an inverter connected typically through a relatively long cable. Theoretically, if the length of the cable is long than quarter of the wave length, the drive PWM pulses appear with a delay at the motor terminals and thus causes reflected pulses. Superposition of a traveling pulse and its refection make the overvoltage ringing at the motor terminal. In Fig. 1, stator windings for only Phase-A is shown, where the winding consists of $N$ number of series coils per phase. Inductance for each coil is represented by $L_{coil}$. Fig. 1 also shows the proposed control branch across the first coil. This branch consists of passive circuit elements controlled by GaN-HEMT switches. The proposed circuit might be placed inside one slot of a large motor due to its compact fabrication.

There might be some technical challenges with this proposed high-payoff circuit, such as the impact on the electromagnetic performance of the motor drives, self-powered GaN drive circuits, and autonomous control of the adaptive impedance in the harsh operating environment inside the electric motors.

Fig. 2 shows the actual line-to-line voltage waveform of a motor drive system measured at the motor terminals. Fig. 2(a) shows one cycle of the voltage waveform measured at the inverter terminal (on the top) and the motor terminal (on the bottom), while Fig. 2(b) shows the zoom-in view of the voltage spikes of the voltage at the motor terminals. As can be seen, the voltage waveform at the motor terminals contains intensive high-frequency voltage spikes with an amplitude twice the DC bus voltage of the drive. Fig. 2(b) shows that the period of these high voltage spikes is about 1.4 μs, corresponding to a high frequency of $714\ kHz$. Notice, the frequency of the ringing is a function of the cable's length and material property.

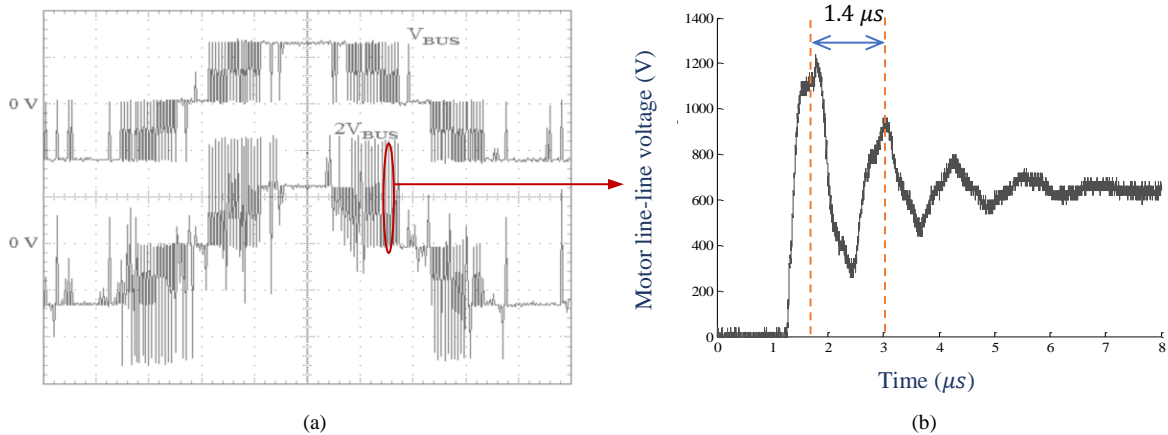

Fig. 2. Measured line-to-line voltages (a) at the inverter (top) and motor terminals (bottom), (b) and the zoom-in view of the line-to-line voltage spike at the motor terminal.

## III. THE PROPOSED CONTROL SCHEME

The circuit shown in Figs. 2 and 3, consisting of a GaN switch and passive elements, provides autonomous impedance matching and mitigate the voltage stress on the motor windings. Remarkably, industry engineers in the motor-drive business have demonstrated that a large portion of the overvoltage due to reflected waves appears across the first coil and the first few turns of stator windings, e.g. [20], [48]. Thus, only the first coil per phase may need to be equipped by the control branch in Fig. 3.

The equivalent impedance of the coil, shown in Fig. 3(a), can be controlled and its impedance value can be changed by controlling the duty cycle of a GaN transistor/switch. The GaN switch is controlled by a self-powered gate driver energized by the stepped-down low voltage circuit from the motor terminal voltage. Both the GaN transistor and ceramic capacitors and/or inductors have a tiny footprint to be located across the first coil and can operate at high temperatures (up to 175°C). Moreover, the GaN transistors are only switched when there are excessive voltage spikes (e.g., 2X DC bus voltage) appearing on the motor input terminals, so the switching and conduction losses are minimal, requiring no additional cooling for the GaN switches.

Suppose the duty cycle of the GaN transistor can be controlled such that the surge impedance of the motor matches the cable surge impedance, then no reflected voltage waveform is generated, and therefore the maximum voltage at the motor terminals does not exceed the dc-bus voltage. A model-reference technique can adjust the motor surge impedance to mitigate the voltage stress across the stator windings in motor-drive systems. In the developed scheme, the impedance matching scheme is updated in real-time based on an adaptation law to ensure that the parallel branch ideally provides a nearly zero-reflection wave at the motor terminals following the desired reference model, thereby ensuring zero-reflection without the need for any cable surge impedance estimation. If the dynamic model of the system can be written as:

$$\frac{d}{dt}Y = g(Y, V_{dc}) \tag{1}$$

The dynamic model for the high-frequency voltage ringing can be expressed by a second-order differential equation as

$$\begin{cases} \frac{d}{dt}X = \begin{bmatrix} a_{11} & a_{12} \\ a_{21} & a_{22} \end{bmatrix} X + bV_{dc}u(t) \\ V_{coil} = CX \end{cases} \tag{2}$$

where, $C$ is a 1-by-2 vector, $V_{dc}u(t)$, is the input step function, and $V_{coil}$, is considered as the output for this system. Regardless of the complexity of the actual circuit/system and the order of the system, a second-order system can be considered as a reference model, representing the desired transient/dynamic behavior as follows:

$$\begin{cases} \frac{d}{dt}X_M = \begin{bmatrix} a_{M11} & a_{M12} \\ a_{M21} & a_{M22} \end{bmatrix} X_M + b_M V_{dc} u(t) \\ V_{coilM} = C_M X_M \end{cases} \tag{3}$$

where, $X_M = [V_{coilM}\ \ i_{HF}]^T$, $C_M = [0\ \ 1]$, $V_{coilM}$ is the reference model output or the desired first coil voltage, and $i_{HF}$ is the high-frequency component of phase current. The general idea is to find the duty ratio, $D$, such that the error between actual measurement, $V_{coil}$, and its desired value, $V_{coilM}$, goes to zero, i.e. $e = V_{coil} - V_{coilM} \to 0$. The coil voltage can be adjusted at any instant by controlling the parallel branch impedance as $Z_{eq} = g(D, f)$, where $V_{coil} = Z_{eq} i_{HF}$.

A reference model for the system with underdamped characteristics can be obtained by setting the diagonal terms to be less than zero, with the off-diagonal terms representing the undamped frequency and having opposite polarities, i.e. $-a_{M11} = -a_{M22}$ and $a_{M21} = -a_{M12}$, where $a_{M11} > 0$. A reference model with critically damped characteristics can be obtained similarly by choosing a diagonal matrix with the diagonal terms being less than zero with the off-diagonal elements set to zero, i.e. $a_{M12}, a_{M21} = 0$. We can investigate both reference models while the parameters can be identified through an optimization algorithm to achieve the minimum power loss in the parallel variable impedance branch. The advantage of this technique is that the actual circuit parameters can be unknown when we force the voltage ringing to be reduced significantly by controlling the duty ratio. The required bandwidth can be achieved when each GaN transistor is switched faster than the $dv/dt$ of switches in the drive. The objective is to adjust the duty ratio, $D$, such that the error between the system output and the reference model output, $e = V_{coil} - V_{coilM} = CX - C_M X_M$, goes to zero over a short time

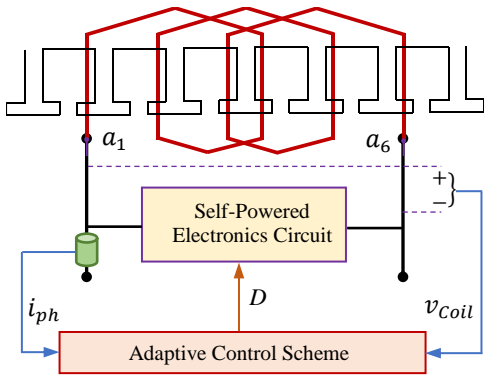
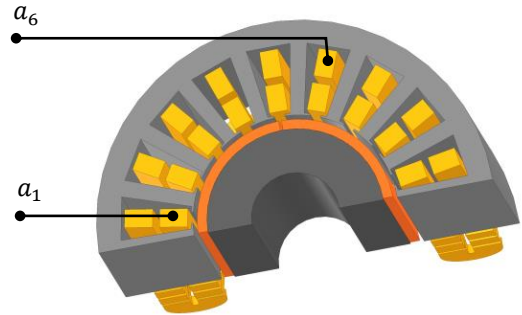

Fig. 3. Stator windings of a typical AC motor, and the location of the power electronic circuit forming a coil with an adaptive surge impedance.

when a traveling pulse hits the motor terminal. The dynamics of the error can be expressed as follows:

$$\dot{e} = C(AX + bV_{dc}) - C_M(A_M X_M + b_M V_{dc}) \quad (4)$$

Adding and subtracting $C_M A_M X$ from the right-hand side of (4), $\dot{e}$ can be restructured as follows:

$$\dot{e} = C_M A_M e + (CA - C_M A_M)X + (Cb - C_M b_M)V_{dc} \quad (5)$$

In the model-reference adaptive theory, the adaptation law is derived by defining a Lyapunov or energy function, $E$, which must be positive definite, i.e. $E > 0$, and its derivative with respect to time must be negative definite, i.e. $\dot{E} < 0$. Herein, a Lyapunov function to extract the equivalent impedance of the RC parallel branch, while the error dynamic in (5) is asymptotically approaching zero can be written in terms of $e$ and $|Z_{eq}|$. The Lyapunov function for (5) can be chosen as follows:

$$E = \frac{1}{2}\big(e^2 + |Z_{eq}| i_{HF}^2\big) \quad (6)$$

The chosen Lyapunov (energy) function in (6) can be seen to be positive definite. The first term is chosen herein to make the error asymptotically approaching zero, while the second term is selected to guarantee that $\dot{E} < 0$, at any time, $t \geq 0$. From (3), $Z_{eq}$ is the function of the duty ratio, $D$, and elements of $A_M$ and $b_M$ are constant values as they are the parameters of the chosen reference model.

## IV. CONCLUSION

The adaptive surge impedance for ac motors presented in this article is a transformative concept to enable the conventional motors much more compatible with the emerging WBG based drives. The benefit of such stator windings with adaptive surge impedance is the mitigation of voltage spikes on the motor terminals with adaptive impedance matching, without sacrificing the drive efficiency and power density. Notice, the conventional solutions are bulky and lossy passive filters that have been used for some motor drive systems, i.e., 15-20% of the drive system weight. The proposed technology eliminates the need for a bulky and lossy $dv/dt$ filter or sinewave filter currently recommended for long-cable motor drive systems. The proposed technology significantly mitigates the insulation stress due to high $dv/dt$ on the stator windings regardless of the cable length or the rise time of the WBG switches. As a result, the overall motor-drive systems can concurrently possess high efficiency, high power density, and high reliability. Additionally, this proposed new approach has great potential to eliminate some other high-frequency transients between WBG drives and ac motors, such as induced voltages and leakage currents.

The proposed solution is a transformative and innovative concept that has not been attempted before for long-cable-fed motor drive systems. Impacts of such adaptive impedance matching circuits that are embedded into the motor windings include the mitigation of common-mode voltage and electromagnetic interference (EMI) issues in the WBG-based drives. Thus, this proposed research is just part of broad new scope to enable motor drives more power-dense, efficient, and reliable.

## V. REFERENCES

[1] A. Adib, *et al.*, "E-mobility – Advancements and challenges," *IEEE Access*, vol. 7, pp. 165226-165240, 2019.
[2] J. Benzaquen, J. He, and B. Mirafzal, "Toward more electric powertrains in aircraft: Technical challenges and advancements," *CES Transactions on Electrical Machines and Systems*, vol. 5, no. 3, pp. 177-193, Sept. 2021.
[3] M. T. Fard, J. He, H. Huang and Y. Cao, "Aircraft distributed electric propulsion technologies-A review," *IEEE Transactions on Transportation Electrification*, vol. 8, no. 4, pp. 4067-4090, Dec. 2022.
[4] Z. Liu, B. Li, F. C. Lee and Q. Li, "High-efficiency high-density critical mode rectifier/inverter for WBG-device-based on-board charger," *IEEE Transactions on Industrial Electronics*, vol. 64, no. 11, pp. 9114-9123, Nov. 2017.
[5] J. Benzaquen and B. Mirafzal, "Seamless dynamics for wild-frequency active rectifiers in more electric aircraft," *IEEE Transactions on Industrial Electronics*, vol. 67, no. 9, pp. 7135-7145, Sep. 2020.
[6] J. Benzaquen, F. Fateh and B. Mirafzal, "On the dynamic performance of variable-frequency AC–DC converters," *IEEE Transactions on Transportation Electrification*, vol. 6, no. 2, pp. 530-539, June 2020.
[7] J. Benzaquen, F. Fateh, M. B. Shadmand and B. Mirafzal, "Performance comparison of active rectifier control schemes in more electric aircraft applications," *IEEE Transactions on Transportation Electrification*, vol. 5, no. 4, pp. 1470-1479, Dec. 2019.
[8] J. Benzaquen, M. B. Shadmand, A. Stonestreet, and B. Mirafzal "A unity power factor active rectifier with optimum space-vector predictive DC voltage control for variable frequency supply suitable for more electric aircraft applications," *IEEE Applied Power Electronics Conference & Exposition* (APEC), March 2018.
[9] J. Benzaquen, M. B. Shadmand, and B. Mirafzal "Ultrafast rectifier for variable-frequency applications," *IEEE Access*, vol. 7, pp. 9903-9911, 2019.
[10] J. Benzaquen and B. Mirafzal, "Smart active rectifier fed by a variable voltage and frequency source," 2021 IEEE Kansas Power and Energy Conference (KPEC), 2021, pp. 1-5.
[11] J. Benzaquen and B. Mirafzal, "An active rectifier fed by a variable-speed generator," 2020 IEEE Applied Power Electronics Conference and Exposition (APEC), New Orleans, LA, USA, 2020, pp. 1691-1696.
[12] J. Benzaquen, A. Adib, F. Fateh and B. Mirafzal, "A model predictive control scheme formulation for active rectifiers with LCL filter," 2019 IEEE Energy Conversion Congress and Exposition (ECCE), Baltimore, MD, USA, 2019, pp. 3758-3763.
[13] J. Benzaquen, M. B. Shadmand, F. Fateh, and B. Mirafzal, "Model reference adaptive one-step-ahead control scheme for active rectifiers in wild frequency applications," IEEE Applied Power Electronics Conference & Exposition (APEC), March 2019.
[14] J. Lamb, A. Singh, and B. Mirafzal, "Rapid implementation of solid-state based converters in power engineering laboratories," IEEE Transactions on Power Systems, vol. 31, no. 4, pp. 2957 – 2964, July. 2016.
[15] I. Husain *et al.*, "Electric Drive Technology Trends, Challenges, and Opportunities for Future Electric Vehicles," *Proceedings of the IEEE*, vol. 109, no. 6, pp. 1039-1059, June 2021.
[16] J. M. Erdman, R. J. Kerkman, D. W. Schlegel and G. L. Skibinski, "Effect of PWM inverters on AC motor bearing currents and shaft voltages," *IEEE Transactions on Industry Applications*, vol. 32, no. 2, pp. 250-259, March-April 1996.
[17] D. Busse, J. Erdman, R. J. Kerkman, D. Schlegel and G. Skibinski, "Bearing currents and their relationship to PWM drives," *IEEE Transactions on Power Electronics*, vol. 12, no. 2, pp. 243-252, March 1997.
[18] R. J. Kerkman, D. Leggate and G. L. Skibinski, "Interaction of drive modulation and cable parameters on AC motor transients," *IEEE Transactions on Industry Applications*, vol. 33, no. 3, pp. 722-731, May-June 1997.
[19] A. Von Jouanne and P. N. Enjeti, "Design considerations for an inverter output filter to mitigate the effects of long motor leads in ASD applications," *IEEE Transactions on Industry Applications*, vol. 33, no. 5, pp. 1138-1145, Sept.-Oct. 1997.
[20] G. L. Skibinski, R. J. Kerkman and D. Schlegel, "EMI emissions of modern PWM AC drives," *IEEE Industry Applications Magazine*, vol. 5, no. 6, pp. 47-80, Nov.-Dec. 1999.
[21] E. J. Bartolucci and B. H. Finke, "Cable design for PWM variable-speed AC drives," in *IEEE Transactions on Industry Applications*, vol. 37, no. 2, pp. 415-422, March-April 2001.

[22] S. Choi, M. S. Haque, M. Tarek, V. Mulpuri, Y. Duan, S. Das, V. Garg, D. Ionel, M. Masrur, B. Mirafzal, and H. Toliyat, "Fault diagnosis techniques for permanent magnet AC machine and drives– A review of current state of the art," *IEEE Transactions on Transportation Electrification*, vol. 4, no. 2, pp. 444 – 463, June 2018.

[23] B. Mirafzal, and N. Demerdash, "On innovative methods of induction motor inter-turn and broken-bar fault diagnostics," *IEEE Transactions on Industry Applications*, vol.42, no. 2, pp.405-414, March/April 2006.

[24] B. Mirafzal, and N. Demerdash, "Induction machine fault diagnosis using the rotor magnetic field space vector orientation," *IEEE Transactions on Industry Applications*, vol.40, no. 2, pp.534-542, March/April 2004.

[25] A. Sayed-Ahmed, C. Yeh, B. Mirafzal, and N. Demerdash, "Analysis of stator winding inter-turn short-circuit faults in poly-phase induction machines for identification of the faulty phase and estimation of the fault severity," *Proc. IEEE Industry Applications Society Conf.*, Tampa, FL, vol.3, Oct. 2006, pp. 1519-1524.

[26] C. Yeh, B. Mirafzal, R. Povinelli, and N. Demerdash, "A Condition monitoring vector database approach for broken bar fault diagnostics of induction machines," *Proc. IEEE International Electric Machines and Drives Conf.*, San Antonio, TX, May 2005, pp.29-34.

[27] A. Sarikhani, B. Mirafzal, and O. Mohammed, "Inter-turn fault diagnosis of PM synchronous generator for variable speed wind applications using floating space vector," *Proc. IEEE Industrial Electronics Society Conf.*, Phoenix, AZ, Nov. 2010, pp. 2628-2633.

[28] B. Mirafzal, and N. Demerdash, "Effects of load magnitude on diagnosing broken bar faults in induction motors using the pendulous oscillation of the rotor magnetic field orientation," *IEEE Transactions on Industry Applications*, vol.41, no. 3, pp.771-783, May/June 2005.

[29] F. Chai, L. Gao, Y. Yu and Y. Liu, "Fault-tolerant control of modular permanent magnet synchronous motor under open-circuit faults," *IEEE Access*, vol. 7, pp. 154008-154017, 2019.

[30] J. Lamb, and B. Mirafzal, "Open-circuit IGBT fault detection and location isolation for cascaded multilevel converters," *IEEE Transactions on Industrial Electronics*, vol. 64, no.6, pp. 4846 - 4856, June 2017.

[31] A. Sayed-Ahmed, B. Mirafzal, and N. Demerdash, "A fault-tolerant technique for Δ-connected AC motor-drives," *IEEE Transactions on Energy Conversion*, vol. 26, no. 2, pp. 646 - 653, June 2011.

[32] B. Mirafzal, "Survey of fault-tolerance techniques for three-phase voltage source inverters," *IEEE Transactions on Industrial Electronics*, vol. 61, no. 10, pp. 5192-5202, Oct. 2014.

[33] B. Poudel, *et al*., "Toward less rare-earth permanent magnet in electric machines: A review," *IEEE Transactions on Magnetics*, vol. 57, no. 9, pp. 1-19, Sept. 2021.

[34] D. Han, C. T. Morris, W. Lee and B. Sarlioglu, "A case study on common mode electromagnetic interference characteristics of GaN HEMT and Si MOSFET power converters for EV/HEVs," *IEEE Transactions on Transportation Electrification*, vol. 3, no. 1, pp. 168-179, March 2017.

[35] A. K. Kaviani, B. Hadley, and B. Mirafzal, "A time-coordination approach for regenerative energy saving in multi-axis motor-drive systems," *IEEE Transactions on Power Electronics*, vol. 27, no. 2, pp. 931 - 941, February 2012.

[36] F. Sadeque, and B. Mirafzal, "Frequency restoration of grid-forming inverters in pulse load and plug-in events," *IEEE Journal of Emerging and Selected Topics in Industrial Electronics*, 2022, doi: 10.1109/JESTIE.2022.3186156.

[37] D. Sharma, F. Sadeque, and B. Mirafzal, "Synchronization of inverters in grid forming mode, *IEEE Access*, vol. 10, pp. 41341-41351, 2022.

[38] M.S. Pilehvar, D. Sharma, and B. Mirafzal, "Forming interphase microgrids in distribution systems using cooperative inverters," *CPSS Transactions on Power Electronics and Applications*, vol. 7, no. 2, pp. 186-195, June 2022.

[39] F. Sadeque, J. Benzaquen, A, Adib, and B. Mirafzal, "Direct phase-angle detection for three-phase inverters in asymmetrical power grids," *IEEE Journal of Emerging and Selected Topics in Power Electronics*, vol. 9, no. 1, pp. 520-528, Feb. 2021.

[40] Adib, F. Fateh, and B. Mirafzal, "Smart inverter stability enhancement in weak grids using adaptive virtual-inductance," *IEEE Transactions on Industry Applications*, vol. 57, no. 1, pp. 814-823, Jan./Feb. 2021.

[41] M.S. Pilehvar, and B. Mirafzal, "Frequency and voltage supports by battery-fed smart inverters in mixed-inertia microgrids," *Electronics* 2020, *9*, 1755.

[42] Mirafzal, and A. Adib, "On grid-interactive smart inverters: features and advancements," *IEEE Access*, vol. 8, pp. 160526-160536, 2020.

[43] T. Hossen, M. Gursoy, and B. Mirafzal, "Self-protective inverters against malicious setpoints using analytical reference models," *IEEE Journal of Emerging and Selected Topics in Industrial Electronics*, vol. 3, no. 4, pp. 871-877, Oct. 2022.

[44] A. Adib, J. Lamb, and B. Mirafzal, "Ancillary services via VSIs in microgrids with maximum dc-bus voltage utilization," *IEEE Transactions on Industry Application*, vol. 55, no. 1, pp. 648-658, Jan.-Feb. 2019.

[45] J. Lamb, B. Mirafzal, and F. Blaabjerg "PWM common mode reference generation for maximizing the linear modulation region of CHB converters in islanded microgrids," *IEEE Transactions on Industrial Electronics*, vol. 65, no. 7, pp. 5250 – 5259, July 2018.

[46] J. Lamb, and B. Mirafzal, "An adaptive SPWM technique for cascaded multilevel converters with time-variant DC sources," *IEEE Transactions on Industry Applications*, vol. 52, no. 5, pp. 4146 – 4155, September/October 2016.

[47] A. Adib, J. Lamb, and B. Mirafzal, "Atypical PWM for maximizing 2L-VSI DC-bus utilization in inverter-based microgrids with ancillary services," Applied Power Electronics Conference & Exposition (APEC), March 2018.

[48] B. Mirafzal, G. Skibinski, R. Tallam, D. Schlegel, and R. Lukaszewski, "Universal induction motor model with low-to-high frequency response characteristics," *IEEE Transactions on Industry Applications*, vol.43, no. 5, pp. 1233 - 1246, September/October 2007.

[49] B. Mirafzal, G. Skibinski, and R. Tallam, "Determination of parameters in the universal induction motor model," *IEEE Transactions on Industry Applications*, vol.45, no. 1, pp. 142 - 151, January/February 2009.

[50] M. Gries, and B. Mirafzal, "Permanent magnet motor-drive frequency response characteristics for transient phenomena and conducted EMI analysis," *Proc. IEEE Applied Power Electronics Conf.*, Austin, TX, February 2008, pp. 1767-1775.

[51] B. Mirafzal, G. Skibinski, and R. Tallam, "A failure mode for PWM inverter-fed AC motors due to the antiresonance phenomenon," *IEEE Transactions on Industry Applications*, vol. 45, no. 5, pp. 1697-1705, September/October 2009.

[52] D. Zhang, J. He and D. Pan, "A megawatt-scale medium-voltage high-efficiency high power density SiC+Si hybrid three-level ANPC inverter for aircraft hybrid-electric propulsion systems," *IEEE Transactions on Industry Applications*, vol. 55, no. 6, pp. 5971-5980, Nov.-Dec. 2019.

[53] J. He *et al*., "Multi-domain design optimization of $dv/dt$ filter for SiC-based three-phase inverters in high-frequency motor-drive applications," *IEEE Transactions on Industry Applications*, vol. 55, no. 5, pp. 5214-5222, Sep.-Oct. 2019.

[54] Y. Zhang, H. Li and F. Z. Peng, "A low-loss compact reflected wave canceller for SiC motor drives," *IEEE Transactions on Power Electronics*, vol. 36, no. 3, pp. 2461-2465, March 2021.

[55] K. K. Yuen, H. S. Chung and V. S. Cheung, "An active low-loss motor terminal filter for overvoltage suppression and common-mode current reduction," *IEEE Transactions on Power Electronics*, vol. 27, no. 7, pp. 3158-3172, July 2012.

[56] Y. Montasser, M. I. Marei and S. H. Jayaram, "Low-power high-voltage power modulator for motor insulation testing," *IEEE Transactions on Industry Applications*, vol. 44, no. 4, pp. 1059-1066, July-Aug. 2008.

[57] B. Mirafzal, Power Electronics in Energy Conversion Systems, New York, USA, McGraw Hill, 2022.